\documentclass[aps,preprint,superscriptaddress]{revtex4}
\usepackage{color}
\usepackage{amsmath,bm}
\usepackage{graphicx}

\begin{document}

\title{Pseudospin rotation and valley mixing in electron scattering at graphene edges}

\author{Changwon Park}
\affiliation{Department of Physics and Astronomy, Seoul National University, Seoul 151-747, Korea}
\author{Heejun Yang}
\affiliation{Semiconductor Devices Lab, Samsung Advanced Institute of Technology, Yongin, Gyeonggi-Do 449-712, Korea}
\author{Andrew J. Mayne}
\author{G\'{e}rald Dujardin}
\affiliation{Laboratoire de Photophysique Mol\'eculaire, CNRS, B\^at. 210, Univ Paris Sud, 91405 Orsay, France}
\author{Sunae Seo}
\affiliation{Semiconductor Devices Lab, Samsung Advanced Institute of Technology, Yongin, Gyeonggi-Do 449-712, Korea}
\author{Young Kuk}
\affiliation{Department of Physics and Astronomy, Seoul National University, Seoul 151-747, Korea}
\author{Jisoon Ihm}
\affiliation{Department of Physics and Astronomy, Seoul National University, Seoul 151-747, Korea}
\author{Gunn Kim}
\email{kimgunn@gmail.com}
\affiliation{Department of Physics, Sejong University, Seoul 143-747, Korea}

\date{\today}
\maketitle

{\bf
In graphene, the pseudospin and the valley flavor arise as new types of quantum degrees of freedom due to the honeycomb lattice
comprising two sublattices ($A$ and $B$) and two inequivalent Dirac points ($K$ and $K'$) in the Brillouin zone, respectively.
Unique electronic properties of graphene result in striking phenomena such as Klein tunnelling,\cite{1,2} Veselago lens,\cite{3} and valley-polarized currents.\cite{4,5,6,7}
Here, we investigate the roles of the pseudospin and the valley in electron scattering at graphene edges and show that they are strongly correlated with charge density modulations 
of short-wavelength oscillations and slowly-decaying beat patterns. 
Theoretical analyses using nearest-neighbor tight-binding methods and first-principles density-functional theory (DFT) calculations 
agree well with our experimental data from the scanning tunneling microscopy (STM). 
We believe that this study will lead to useful application of graphene to ``valleytronics" and ``pseudospintronics".
}
\\

A variety of edge properties\cite{8,9,10,11} of graphene and graphene nanoribbons\cite{12} have been investigated and interference images using the STM were also reported before.\cite{13} 
However, electron scattering behaviours at graphene edges have not been well understood yet. 
A conventional metal with a terrace and a step can be thought as accomodating a two-dimensional (2D) free electron gas with a hard wall edge and the standing wave formed at the edge can be analytically solved. 
This behaviour was directly observed at the steps of Au(111) and Cu(111) surfaces by scanning tunneling microscopy (STM).\cite{14,15,16} 
Now, a question arises as to whether the graphene edge has a similar standing wave pattern to conventional metals, 
and how two sublattices and two inequivalent valleys in graphene affects the scattering and the standing wave formation.
We show below that the interference pattern at the graphene edge is quantitatively understood in terms of intra- and intervalley scattering processes at the graphene edge 
and that local defects at the edge considerably change the interference pattern in certain cases.

Due to the crystal momentum conservation along the edge, available backscattering channels are limited to the Bloch states of the same wavevector component in the edge direction ($k_y$) as the incident wave. 
The two distinct Fermi circles around $K$ and $K'$ valleys in the doped graphene are folded back onto the rectangular Brillouin zone (equivalent to the hexagonal first Brillouin zone) 
in the armchair and the zigzag edge cases, as depicted in Figs. 1\textbf{a} and 1\textbf{b}, respectively. For a given wavevector $k_y$, there are two intersecting points for each Fermi circle, 
corresponding to left and right propagating states. Figures 1\textbf{c} and 1\textbf{d} show two possible (intravalley and intervalley) backscattering channels 
at the armchair edge whereas only one intravalley scattering channel is allowed at the zigzag edge.\cite{17} In the nearest-neighbor tight-binding method, the scattering wave at graphene edges can be uniquely determined by boundary matching conditions. As shown in the left panels of Fig. 2, we choose three representative model structures\cite{18} for the armchair edge termination that have been observed 
in the transmission electron microscopy image\cite{10} as well as another structure of the zigzag edge termination.
The resulting reflection probabilities and electronic density profile are calculated with appropriate hopping parameters from the first-principles Wannier function analysis (See Supplementary Information) and the Fermi energy ($E_F$) is chosen to be 0.3 eV above the Dirac point ($E_D$). 
Note that the epitaxial graphene on the SiC(0001) surface is n-doped and typically $E_F - E_D$ $\approx$ 0.3 $-$ 0.4 eV.\cite{19}

Considering the crystal momentum conservation at the armchair edge, intervalley and intravalley scatterings can in principle be allowed as mentioned above (See Fig. 1\textbf{c}). 
However, when the boundary matching conditions are applied (See Supplementary Information), the actual scattering at the armchair edge is an entirely intervalley process for all incident angles,\cite{20} 
and the pseudospin of the incident wave is identical to that of the scattering wave as can be inferred from Fig. 1\textbf{a}. 
In other words, the pseudospin is invariant throughout the scattering process at the ideal armchair edge. 
(In the pseudospin formalism, $A$ and $B$ sublattice components of the wavefunctions are described as up and down ``spins".) 
By integrating all scattering waves on the Fermi surface, 
we construct laterally averaged electronic density profile in the real space (in the right panel of Fig. 2). 
For the armchair edge, the electronic density profile can be written as

\begin{eqnarray}
&\int_{-\frac{\pi}{2}}^{\frac{\pi}{2}}  |\psi_K^{\theta ,in} (x)+\psi_{K'}^{\theta ,out}(x)|^2 
+|\psi_{K'}^{\theta ,in}(x)+\psi_K^{\theta ,out}(x)|^2\,d\theta \nonumber \\
=&\int_{-\frac{\pi}{2}}^{\frac{\pi}{2}}|\exp(i(-K_0-k_x)x)-\exp(i(K_0+k_x)x))|^2 d\theta\nonumber \\
+ &\int_{-\frac{\pi}{2}}^{\frac{\pi}{2}}|\exp(i(-K_0+k_x)x)-\exp(i(K_0-k_x)x))|^2 d\theta\nonumber \\
&\propto 1-\cos (2K_0x)J_0(2k_Rx),
\end{eqnarray}
where $x$ is the distance from the edge, $ \psi_{K(K')}^{\theta ,in(out)} (x)$ is the incident (scattering) planewave on the $K$($K'$)-valley Fermi surface with the angle $\theta$, 
$K_0$ the length of $\overline{\Gamma K}$ in $k$-space, $k_R$ the radius of the Fermi circle and  $J_0$ the zero-order Bessel function.

Very interestingly, the armchair edge has intervalley scattering which is not shared by the 2D free electron gas. 
In the electronic density profile near the armchair edge, a short-period oscillation and a slowly decaying beat pattern are observed.
First, a short-period oscillation comes from the intervalley scattering, and its wavelength is approximately ~1.85 \AA, 
i.e., $3a/4$ where $a$ is the length of graphene lattice vector (0.246 nm). 
Therefore, this phenomenon is a unique property of materials with multiple Fermi circles.
Second, the slowly-decaying oscillatory behaviour has the same origin as the one observed on the terrace of conventional metal.\cite{14}    
Its characteristic period is a few nanometres. 
The Bessel-function-type envelope pattern is in general caused by quantum interference between incident and scattered electrons in 2D metals at the straight edge termination. 
Finally,  a beat is an interference between two waves of slightly different wave vectors. 
As depicted in Fig. 1(c), $|\vec{k}_{2} - \vec{k}_{1}|$ ($K$ to $K'$ scattering) is slightly different from $|\vec{k}_{3} - \vec{k}_{4}|$ ($K'$ to $K$ scattering), and these two intervalley scattering events result in the beat. 
Since the beat is originated from the finite radius of the Fermi circle (\textit{i.e.}, nonzero $|E_F-E_D|$), its period changes as the chemical potential (or the doping level) of graphene is changed.

Although the dangling bond causes the increase in the hopping parameter at the edge, it hardly changes the proportion of two backscattering channels, 
and the electron density profile of the armchair edge with the dangling bonds (Fig. 2\textbf{b}) is almost indistinguishable from the hydrogen-passivated armchair edge (Fig. 2\textbf{a}). 
For the pentagonal reconstruction (called a 5-6 configuration) at the armchair edge (Fig. 2\textbf{c}), both the intervalley and the intravalley scatterings take place for the obliquely incident wave
because of different boundary matching conditions from the ideal edge, 
which means that the orientation of pseudospin now changes. Although this mixing of intervalley and intravalley scattering channels slightly modifies
the charge modulation pattern, overall features of the electronic density profile in the right panel of Fig. 2\textbf{c} 
looks similar to those of the hydrogen-passivated armchair edge. 
In particular, we note that the node-like structure survives here. 
The reason is that the small-angle incident waves have large weights when forming a standing wave near the edge 
and are reflected dominantly via intervalley scattering for the armchair edge regardless of the details of edge termination.

At the zigzag edge, on the other hand, only the intravalley scattering is allowed which shows a standing wave quite similar to the conventional metal surface.
The electronic density profile of the zigzag edge is given by
\begin{eqnarray}
&\int_{-\frac{\pi}{2}}^{\frac{\pi}{2}}  |\psi_K^{\theta ,in} (x)+\psi_K^{\theta ,out}(x)|^2 
+|\psi_{K'}^{\theta ,in}(x)+\psi_{K'}^{\theta ,out}(x)|^2 d\theta \nonumber \\
=&2\int_{-\frac{\pi}{2}}^{\frac{\pi}{2}}|\exp(-ik_xx)-\exp(ik_xx))|^2 d\theta\nonumber \\
&\propto 1-J_0(2k_Rx).
\end{eqnarray}
Here, to reveal the characteristic feature of the scattering unambiguously, the contribution of the edge states at the zigzag edge is not included in the electronic density profile. 
With the intravalley scattering at the zigzag edge, the pseudospin changes its orientation. 
Since the scattering is confined to the same valley, the pseudospin follows the change of the crystal momentum by scattering.
This is the sharply constrasted behaviour of pseudospins at the armchair edge discussed above.

For a low-energy scattering, the diameter of the Fermi circle in graphene is small. 
Specifically, the Fermi circle diameter of slightly n-doped epitaxial graphene on the SiC(0001) surface is about 0.1 \AA$^{-1}$ 
so that the intravalley scattering makes a long-period charge modulation pattern of a characteristic period of $\sim$3 nm. 
Figure 3\textbf{a} is the real space topographic image and Fourier-filtered line profile of the experimental STM data of the armchair edge (see the Methods section). 
By Fourier-filtering, it is possible to suppress other scattering effects from the substrate or from some unidentified defects which may exist 
in graphene.\cite{21} The DFT-simulated STM image for the hydrogen-passivated armchair edge in Fig. 3\textbf{b} agrees excellently with the experimetal result (Fig. 3\textbf{a}).
We have also simulated the armchair edge with the 5-6 configuration\cite{18} and demonstrated that the atomic scale node-like structure in the $y$-axis direction also occurs, as shown in Fig. 3\textbf{c}. 
These are contrasted to the case of the zigzag edge in Fig. 3\textbf{d} where no node-like structure but a periodic modulation is observed.

Next, we move on to scattering properties of graphene with super-periodic edge structures with atomic-scale defects in Fig. 4.
In reality, graphene sheets usually have such complex edge structures.
The super-periodic armchair edges have no other propagating backscattering channels than the original intervalley and intravalley ones, 
if the radius of the Fermi surface is sufficiently small. 
A double-vacancy case and a single-adatom case for different supercell periods are considered.  
The solid (dashed) lines in Fig. 4\textbf{a} and 4\textbf{b} are reflection probabilities of the armchair edge with relatively high (low) defect density,  
and the reflection probability is expected to approach asymptotically to the perfect armchair case in the low density defect limit.
The overall shape of the reflection probability is quite insensitive to the hopping parameters of edge sites and the incident energy.

When vacancies are introduced in the zigzag edge, on the other hand, the scattering channels change drastically depending on the period of the edge. 
When the period becomes three, the intervalley channel opens because both $K$ and $K'$ fall on the $\Gamma$-point in the folded Brillouin zone, and it becomes a dominant backscattering channel. 
If a carbon atom is attached to the zigzag edge, the scattering becomes rather complicated. 
Though the reflection probability converges to the perfect zigzag edge limit at a low density of adatoms, 
a high density of adatom makes the reflection probability strongly depend on hopping parameters and the incident electron energy. 
It is also found that there exists an abrupt suppression of one channel around the incident angle of $90^{\circ}$  in the super-periodic zigzag edge, 
which is due to the asymmetric trigonal warping\cite{22} between $K$ and $K'$ valleys. 
This asymmetry renders only the intravalley scattering channel available\cite{23} for a certain $k_y$ range for the zigzag edge, 
and that property was exploited for generating valley-polarized currents in the former theoretical work.\cite{24} 
Such a suppression is negligible at the armchair edge because of the mirror-symmetric pattern of trigonal warping with respect to the $k_y$-axis. 
(See the Fig. S2 of Supplementary Information for the warping of $K$ and $K'$ valleys for the armchair and zigzag edges.)
More detailed analysis and understanding of the scattering at realistic edges with defects certainly require enormously more work in the future.

\section*{Methods}

{\bf Theory} We used the density functional theory\cite{25,26} calculations and the tight-binding boundary matching methods
to obtain the real-space STM images and reflection probabilities at graphene edges.
The Perdew-Burke-Ernzerhof functional form\cite{27} was adopted in the generalized gradient approximation and the ionic potentials were described
by projector-augmented waves\cite{28} (PAW) implemented in the Vienna Ab initio Simulation Package (VASP).\cite{29}
The plane wave basis with the kinetic energy cutoff at 300 eV was employed for describing wavefunctions, and graphene edges were fully relaxed
until the force on each atom was within 0.02 eV/\AA. Tight-binding parameters were obtained from maximally localized Wanneir function analysis using wannier90.\cite{30}

{\bf Experiment} The graphene was grown epitaxially on the silicon face of a highly n-doped 6H-SiC(0001) by thermal desorption of silicon at high temperature.
To obtain high quality graphene and its edge structure, the pressure during the thermal desorption was kept below $3 \times10^{-9}$ Torr.
The STM images were obtained at 300 K in ultrahigh-vacuum with an Omicron instrument. In monolayer graphene, all hexagons were clearly seen
and we could easily confirm the crystallographic direction. 
In order to filter out other effects such as the substrate structure and the graphene honeycomb lattice structure coexisting in the STM image, we applied the Fourier Filtering in the WSxM software.

\section*{Acknowledgements}
We acknowledge the support of the Core Competence Enhancement Program (2E21580) of KIST through the Hybrid Computational Science Laboratory and the Korean Government MEST Basic Research Grant No. KRF-2006-341-C000015 (C. P. and J. I.), and 
%the Basic Science Research Program through the NRF of Korea funded by the MEST 
Grant No. 2010-0007805
(G.K.). Computations were performed through the support of KISTI. 

\section*{Author contributions}
C.P. and H.Y. carried out theoretical and experimentail researches, respectively, and wrote the paper.
Y.K., A.M., S.S. and G.D. contributed to analysing and interpreting the data.
G.K. and J.I. designed research and edited the paper.

\begin{figure}
\includegraphics[width=0.8\columnwidth]{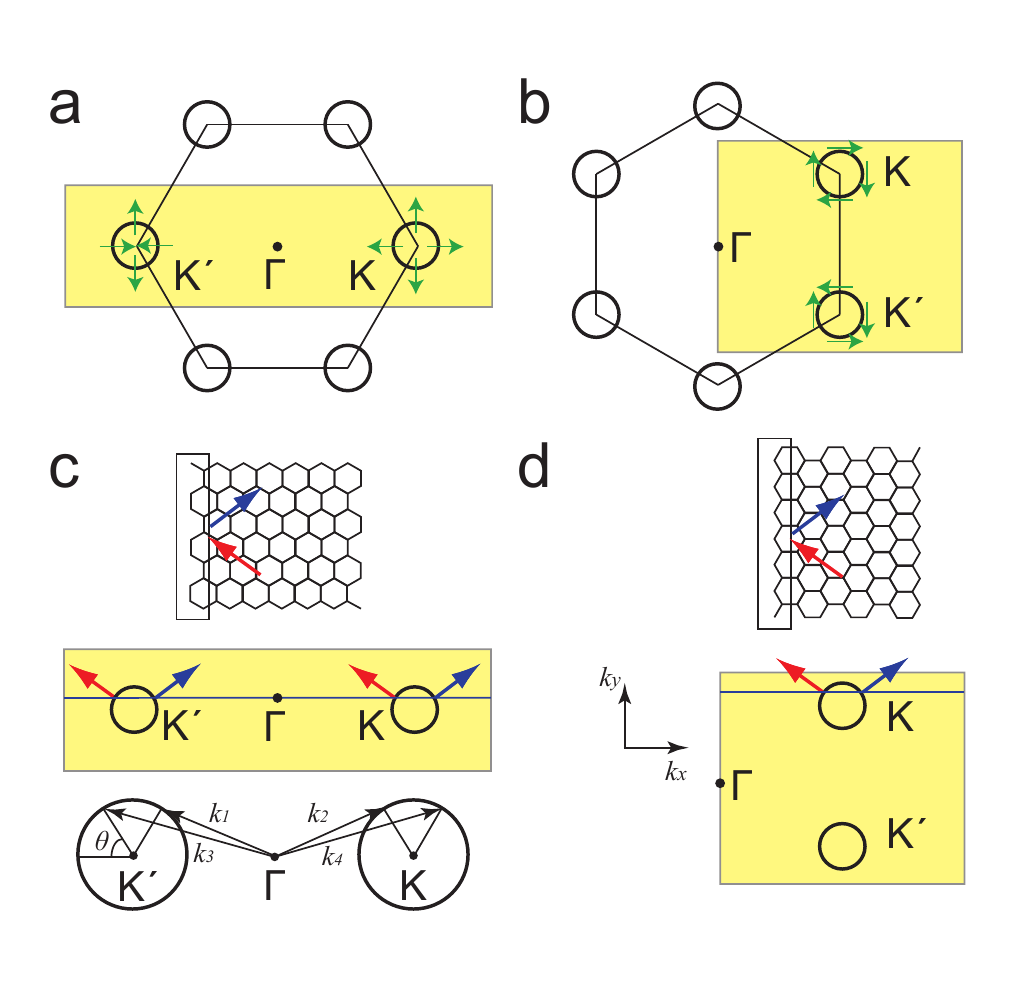}
\caption{\textbf{Scattering channels at the graphene edge.} \textbf{a},\textbf{b}, Schematic Fermi surfaces and pseudospin fields of the armchair edge (\textbf{a}) and the zigzag edge (\textbf{b}). 
The Brillouin zone is transformed to a rectangle in each case, for convenience in taking into account the translational symmetry
in the $y$-direction alone by the presence of the edge. Green arrows stand for orientations of pseudospin fields.
\textbf{c},\textbf{d}, Available scattering channels of a given incident wave for the armchair edge (\textbf{c}) and the zigzag edge (\textbf{d}).
Red arrows indicate the incident direction and blue arrows the scattering direction, respectively. 
In \textbf{d}, the incident wave is assumed to be $K$-valley polarized.
In \textbf{c}, $\theta$ and $k$ denote the incident angle and the Bloch wavevector, respectively. 
The size of the Fermi circle is exaggerated for visual clarity.}
\label{fig1}
\end{figure}

\begin{figure*}
\includegraphics[width=0.7\columnwidth]{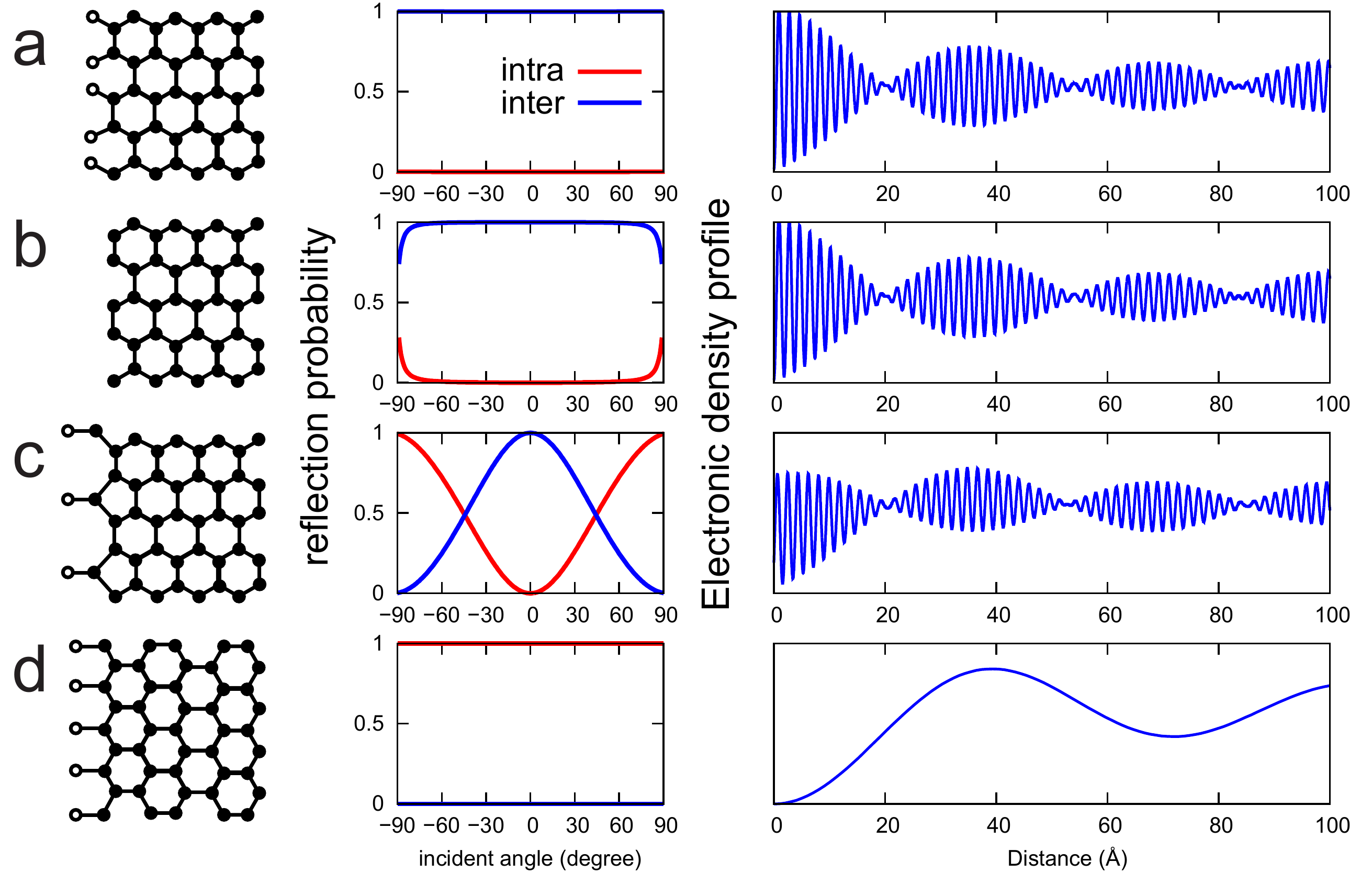}
\caption{\textbf{Reflection probability of the $K$-valley polarized incident wave and the laterally averaged electronic density profile for different edge geometries.} 
\textbf{a},\textbf{b},\textbf{c},\textbf{d}, Reflection probabilities of the $K$-valley polarized incident wave (central panels) and laterally averaged electronic density profile from both valleys 
(right panels) at the hydrogen-passivated armchair edge (\textbf{a}), the dangling armchair edge (\textbf{b}), the reconstructed armchair edge (\textbf{c}), 
and the hydrogen-passivated zigzag edge (\textbf{d}), respectively. The Bloch cell-periodic part is ignored for clarity.
The single hopping parameter of $-2.88$ eV is sufficient to describe the hydrogen-terminated armchair graphene edge.
At the edge of the dangling bond, however, our Wannier function analysis demonstrates that hopping parameters are
increased by 30\% along the edge.
At the edge of the pentagonal reconstruction structure in \textbf{c}, on the other hand, hopping parameters are reduced by 30 $-$ 50\%.
} 
\label{fig2}
\end{figure*}

\begin{figure}
\includegraphics[width=0.7\columnwidth]{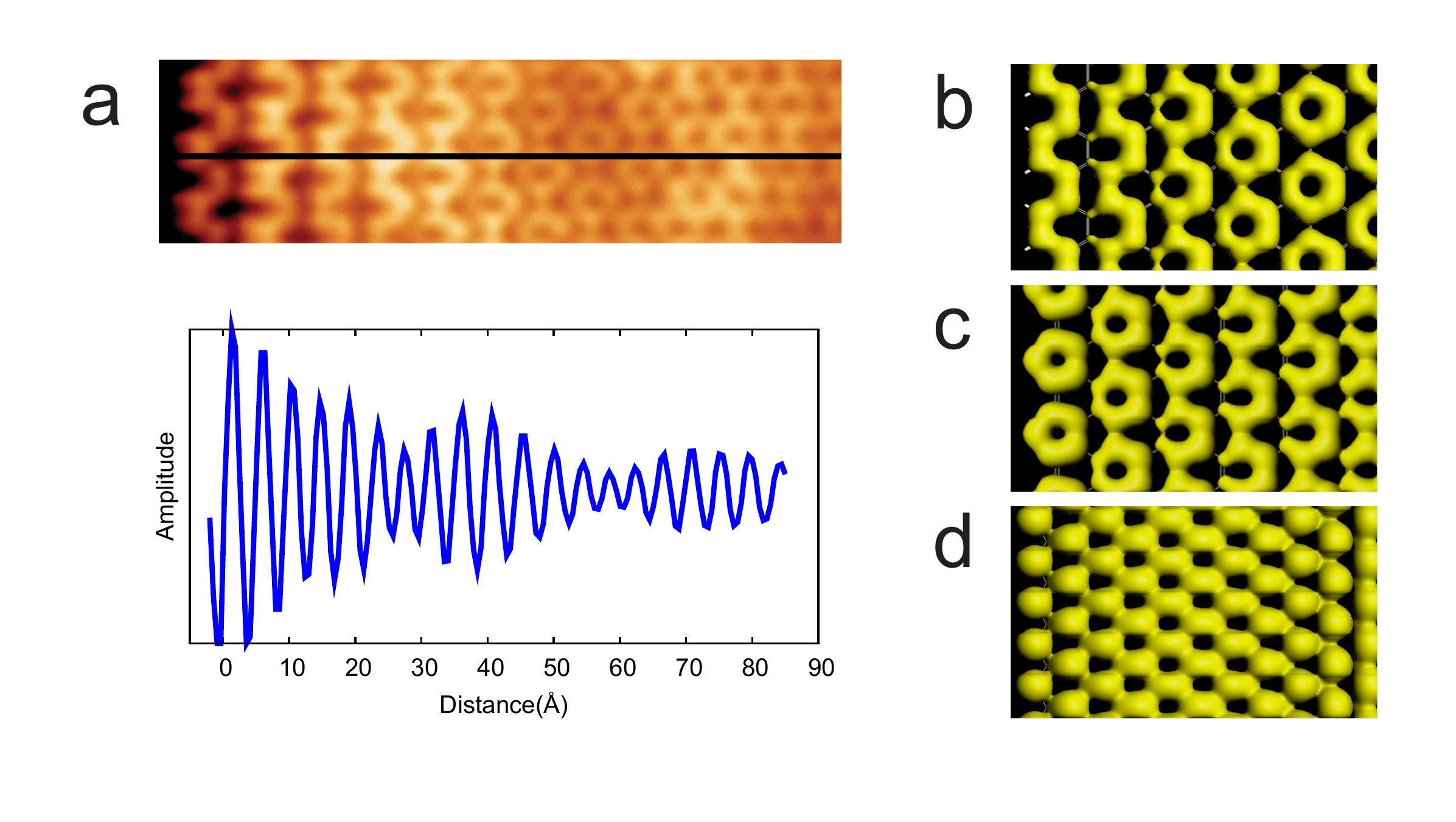}
\caption{\textbf{STM topographic image of the graphene edge.} 
\textbf{a}, Experimentally observed STM topographic image of the graphene armchair edge and its Fourier-filtered profile along the black straight line in the image. \textbf{b},\textbf{c},\textbf{d} DFT-simulated STM images for the hydrogen-passivated armchair edge (\textbf{b}), the 5-6 configuration edge (\textbf{c}), and the zigzag edge (\textbf{d}).
The intervalley scattering manifests itself in a clear node-like structure of the charge density in the $y$-axis direction in \textbf{b} and \textbf{c}, and the DFT-simulation in \textbf{b} is in excellent agreement with the experimental image in \textbf{a} near the edge.
} 
\label{fig3}
\end{figure}

\begin{figure}
\includegraphics[width=0.7\columnwidth]{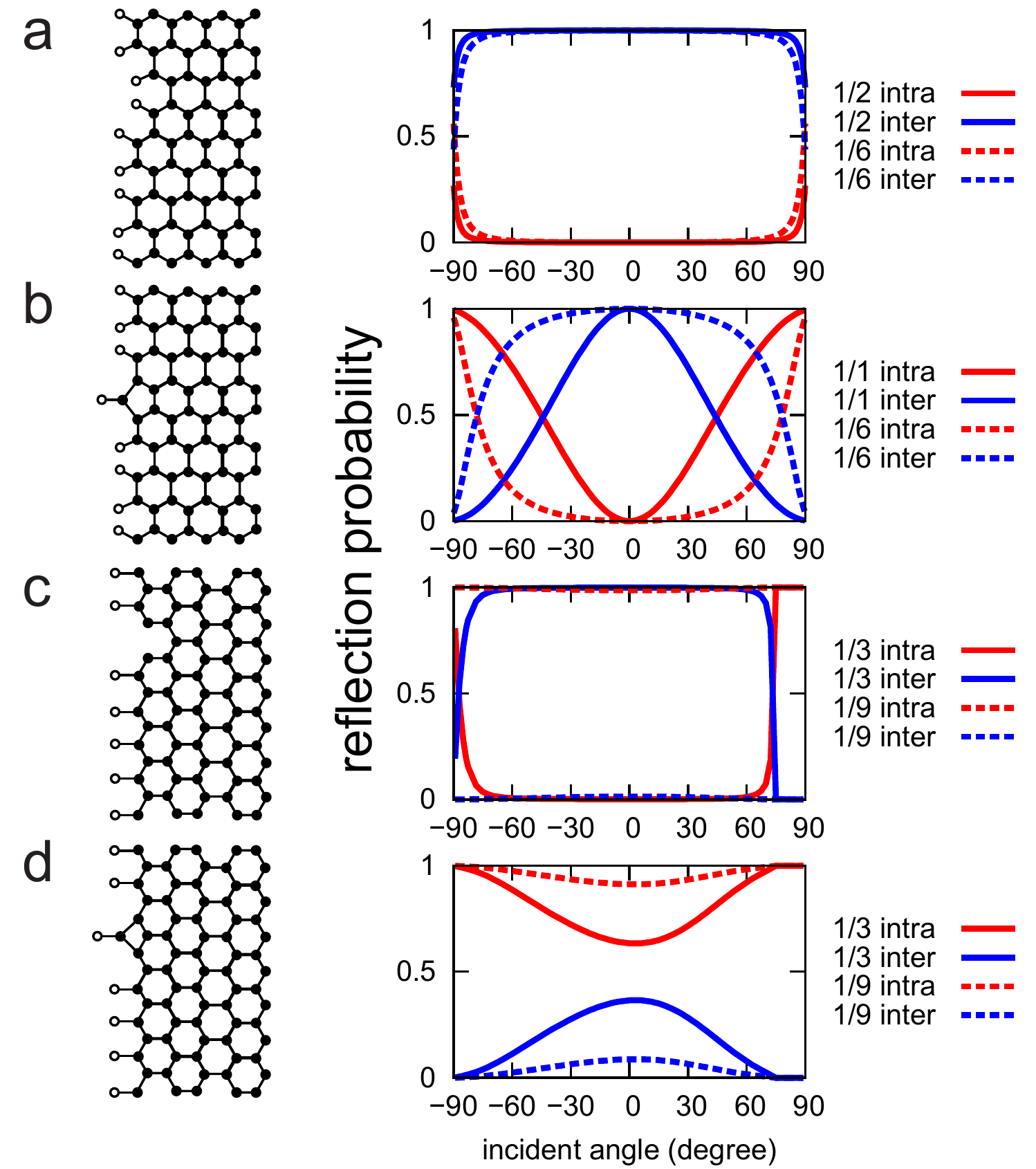}
\caption{\textbf{Reflection probability of the $K$-valley polarized incident waves.} \textbf{a},\textbf{b},\textbf{c},\textbf{d}, Reflection probability of the $K$-valley polarized incident wave of double vacancies at the armchair edge (\textbf{a}), adatoms at the armchair edge (\textbf{b}), single vacancies at the zigzag edge (\textbf{c}), and adatoms at the zigzag edge (\textbf{d}). Fractional numbers indicate the number of defects per unitcell along the edge.
For the super-periodic zigzag edge with defects, the intervalley scattering channel is open when the period is a multiple of three,
because $K$ and $K'$ points coincide with the $\Gamma$-point when the zigzag Brillouin zone is folded three times.
Therefore, both intervalley and intravalley scatterings are allowed for the 3$m$ ($m$: an integer) period. Since only the intravalley scattering is allowed at the (3$m$+1) and (3$m$+2) period edges, the plots of
reflection probabilities for these cases are not presented.
} 
\label{fig4}
\end{figure}

\end{document}